# Femto/nano-second switchable passively mode-locked fiber laser with analytic modeling by cubic-quintic Ginzburg-Landau equation

Wen-Hsuan Kuan,[1,2] Li-Ting Kao, Jing-Yun Wang, and Kuei-Huei Lin[1,3]

[1]*Department of Applied Physics and Chemistry, University of Taipei, 1, Ai-Guo West Road, Taipei, Taiwan*
[2]*e-mail: wenhsuan.kuan@gmail.com*
[3]*e-mail: khlin@utaipei.edu.tw*

**We report a passively mode-locked erbium-doped fiber laser with pulsewidths switchable from 473 fs to 76.8 ns, where the fundamental mode-locking, noise-like pulse, nanosecond mode-locking, and dual-width mode-locking are obtained by adjusting a polarization controller. Co-existence of femto- and nano-second pulses in dual-width mode-locking is attributed to the gain balancing. Analytic modeling of the fiber laser with cubic-quintic Ginzburg-Landau equation is presented, in which the pulsewidths are calculated as functions of dispersion, saturable absorption, and self-phase modulation. The generation of nanosecond pulses is attributed to weakened intracavity pulse-shortening strength and reduced effective self-phase modulation in the laser cavity.** © 2017 Optical Society of America

http://dx.doi.org/10.1364/OL.99.099999

Pulsed fiber lasers have found many emerging applications in laser science, optical communication, biomedical research, and industrial technology, etc. To date, the most widely available gain media for fiber lasers are erbium-doped fiber (EDF) and ytterbium-doped fiber (YDF), which have been used in the wavelength regimes of 1550 nm and 1060 nm, respectively [1–5]. In order to generate ultrashort pulses in fiber lasers, passive mode-locking has been widely reported based on several types of mechanism. One type of passive mode-locking exploits the Kerr nonlinearity of optical fibers to achieve nonlinear polarization evolution (NPE) in the laser cavity [2]. Another type is based on fast saturable absorbers, such as multiple-quantum-well semiconductor devices [1] or nanomaterials [6–12]. It is also feasible to obtain ultrashort pulses in a figure-8 fiber laser with either a nonlinear-optical loop mirror (NOLM) or a nonlinear amplifying loop mirror (NALM) [13,14], or in a fiber laser with a hybrid mode-locking mechanism [3].

Many phenomena such as fundamental mode-locking (FML), pulse splitting, harmonic mode-locking (HML), and Q-switched mode-locking (QML) have been experimentally observed in fiber lasers. For cavity lengths of several to tens of meters, the output pulsewidths of passively mode-locked erbium-doped fiber lasers (EDFLs) are typically tens of femtoseconds to several picoseconds. To obtain nanosecond mode-locked pulses, the cavity length should be extended to hundreds of meters or even longer. Xu et al. demonstrated a mode-locked nanosecond erbium-doped fiber laser, in which the pulsewidth can be varied from 3 ns to 20 ns at a repetition rate between 1.54 MHz and 200 kHz by changing the cavity length [15]. In addition to the ordinary mode-locking states, it was found that fiber lasers can support the generation of noise-like pulses, generating a trail of quasi-stable packets of ultrashort pulses composed of picosecond wavepackets with randomly distributed femtosecond inner pulses [16,17].

In this paper, we present an erbium-doped fiber laser capable of discrete pulsewidth tuning from the femtosecond to the nanosecond regime. Various laser operation states including continuous-wave (CW), fundamental mode-locking, noise-like pulse, nanosecond mode-locking (NML), and dual-width mode-locking (DML) are obtained by simply adjusting an intracavity polarization controller. We will present an analytic modeling of mode-locked fiber lasers by cubic-quintic Ginzburg-Landau equation (CQGLE), in which the pulsewidths will be expressed as functions of intracavity dispersion, saturable absorption, and self-phase modulation. The modeling is then applied to the calculation of femto/nano-second passively mode-locked EDFL.

Figure 1 shows the experimental setup of our ring-cavity EDFL. The output port of a C-band erbium-doped fiber amplifier (EDFA) is connected to its input port through a section of single-mode fiber (SMF), an in-line polarizer, a polarization controller, and a 50/50 output coupler. The length of the erbium-doped fiber is 14 m and the cavity length of the EDFL is about 56 m. The combination of polarizer and polarization controller works effectively as an intensity-dependent loss mechanism, which results from the nonlinear polarization evolution of the intracavity pulses. The laser diode (LD) pump power of the EDFA is set to 127 mW. All the fiber components in the laser cavity are linked via single-mode fibers with FC connectors. While manually adjusting the polarization controller, the laser operating states are monitored by using a high-speed InGaAs photodetector and a digital oscilloscope. The EDFL output is characterized by a sampling oscilloscope, an optical spectrum analyzer, a radio-frequency (RF) spectrum analyzer, and an optical power meter. A noncollinear autocorrelator (Femtochrome FR-103XL) was used to measure the width of the mode-locked pulses.

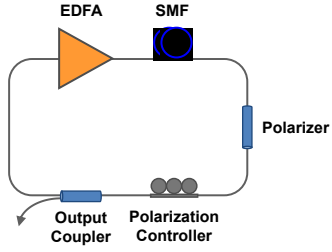

Fig. 1. Experimental setup of the ring-cavity erbium-doped fiber laser.

By carefully adjusting the polarization controller, various operation states including continuous-wave, fundamental mode-locking, pulse-splitting, harmonic mode-locking, and nanosecond mode-locking have been observed in the EDFL. Figure 2 shows the oscilloscope traces as well as the corresponding optical spectra and RF spectra for fundamental mode-locking [Fig. 2(a)–(c)], nanosecond mode-locking [Fig. 2(d)–(f)], and dual-width mode-locking states [Fig. 2(g)–(i)]. The average powers are 6.065 mW, 5.918 mW, and 5.274 mW, respectively. As shown in Fig. 2(a), the pulse repetition rate of fundamental mode-locking is 3.50 MHz, with a peak wavelength of 1562.1 nm and a 3-dB bandwidth of 13.6 nm [Fig. 2(b)].

An interesting phenomenon of nanosecond mode-locking is observed in this EDFL, of which the pulsewidth is about tens of nanoseconds. For nanosecond mode-locking, a pulsewidth of 76.8 ns has been observed [Fig. 2(d)]; however, the peak wavelength shifts to 1532.6 nm, with a 3-dB bandwidth of 1.4 nm [Fig. 2(e)]. We observed that two subsets of mode-locked pulse trains can exist simultaneously in this ring-cavity EDFL. As shown in Fig. 2(g), the pulse trains of broad and short pulses are interleaved with tunable delay, and the spectrum of this dual-width mode-locking state shows two spectral bands peaked at 1532.1 nm and 1567.8 nm, with 3-dB bandwidths of 0.5 nm and 14.2 nm, respectively. Using a tunable band-pass optical filter, the 1532.1-nm and 1567.8-nm spectral bands are confirmed to correspond to the broad and short pulses, respectively. By analyzing the DML spectrum in Fig. 2(h), the power ratio of broad and short pulses is found to be 1:1.04. Therefore, the generation of dual-width pulses is attributed to the balancing of net gain for the two spectral bands, which is achieved by carefully adjusting the polarization controller.

Figures 3(a) and (b) show the autocorrelation trace and curve-fitting of the FML pulses in Fig. 2(a). We observed that the autocorrelation trace consists of a broad base and a narrow center peak, which is the characteristic of noise-like pulses. Using Gaussian fitting, the pulsewidth is calculated to be 714 fs. For the dual-width mode-locking state, the width of the broad pulses in Fig.

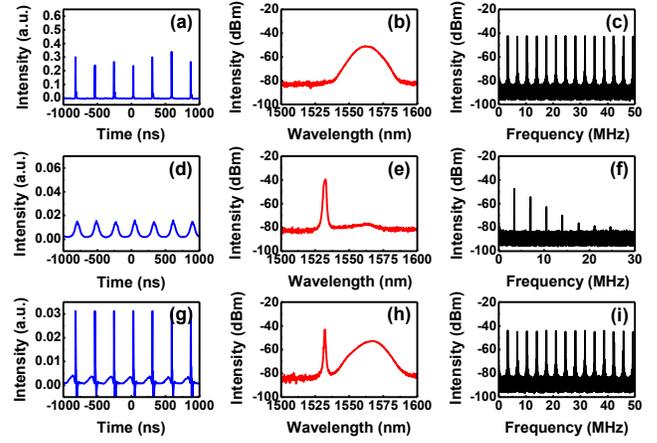

Fig. 2. The oscilloscope traces, optical spectra, and RF spectra for fundamental mode-locking [(a)-(c)], nanosecond mode-locking [(d)-(f)], and dual-width mode-locking states [(g)-(i)].

2(g) is measured as being 74.1 ns, while the autocorrelation trace of the short pulses in Fig. 2(g) shows noise-like pulses of 688 fs [Fig. 3(c)-(d)].

At first glance, the generation of passively mode-locked pulses of several tens of nanoseconds with a cavity length of only 56 m is not technically available since the net intracavity dispersion is not able to support the formation of such ultra-long pulses in the EDFL. In addition, there is not any large-dispersion material such as graphene used in the laser cavity. Therefore, some mechanism should be accounted for these ultra-long laser pulses.

Considering the configuration of this fiber laser, we notice that the pulsewidth and stability of passively mode-locked lasers depend on both the intracavity dispersion and optical nonlinearity which includes self-phase modulation (SPM) and saturable absorption (SA) effects. Since the effective SA is induced by the NPE cavity configuration, the SA as well as the mode-locking strength can be finely tuned via adjusting the intracavity polarization controller. In contrast to the saturable materials with monotonically increasing transmission versus incident intensity, the transmission of NPE SA is a slowly-decaying sinusoidal function with multiple peaks [18]. Therefore, higher-order nonlinear terms should be introduced to simulate the NPE SA. Based on previous analysis of multi-parametric control [19], we suggest that the generation of nanosecond pulses is attributed to the weakened intracavity pulse-shortening strength provided by the artificial SA of the NPE configuration.

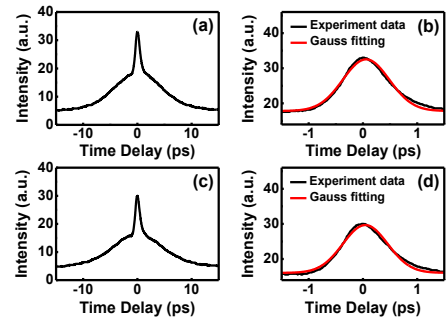

Fig. 3. The autocorrelation traces and curve-fitting for the pulses in pure FML state [(a)-(b)] and the short pulses in DML state [(c)-(d)].

In order to explore the physical mechanism behind the experimental observations, we solve for the cubic-quintic Ginzburg-Landau equation [2]:

$$\left[-j\psi - (\alpha + jx) + g\left(1 + \frac{1}{\Omega}\frac{d^2}{dt^2}\right) + jD\frac{d^2}{dt^2} + (\gamma - j\delta)|a|^2 + \eta|a|^4\right]a = 0, \quad (1)$$

where $a$ is the complex envelope of the electric field of angular frequency $\omega_0$, $\psi$ is the phase shift induced by the carrier frequency shift within an effective optical length, $\alpha$ and $x$ are the linear loss and phase shift of the complex amplitude, $g$ is the gain coefficient, $\Omega$ is the gain bandwidth, $D$ is the dispersion parameter, $\gamma$ is the cubic SA parameter, $\delta$ is the SPM parameter, and $\eta$ is the quintic SA parameter. A particular solution of the cubic-quintic Ginzburg-Landau equation is

$$a = \sqrt{\frac{A}{\cosh\left(\frac{t}{\tau}\right) + B}} \exp\left[-j\frac{\beta}{2}\ln\left[\cosh\left(\frac{t}{\tau}\right) + B\right] + j\theta z\right], \quad (2)$$

in which the pulse is described by the pulse amplitude $A$, width $\tau$, shaping parameter $B$, and the chirp parameter $\beta$. With this ansatz, the CQGLE can be recast into six simultaneous equations, which are solved analytically to obtain the pulsewidths as functions of $D$, $\gamma$, $\delta$, and $\eta$.

In Fig. 4, we demonstrate the dependences of the pulsewidth on the dispersion and SPM for both NML and FML pulses. In Fig. 5, we further show the pulsewidth as functions of cubic and quintic SA. In both figures, the optical parameters of the calculated NML pulses are with central wavelength $\lambda_0$ = 1532.6 nm, bandwidth $\Delta\lambda$ = 1.4 nm, EDFA input power $P_i$ = 2 mW, EDFA output power $P_o$ = 49.32 mW, gain $G$ = 13.93 dB, loss $\alpha$ = 47.32 mW, and repetition rate $R$ = 3.5 MHz, which give to average pulse energy $E_p$ = 7.33 nJ, bandwidth $\Omega$ = 1.12 THz, $g$ = 6.95, $\delta$ = 9.96816 × $10^{-8}$ W$^{-1}$, $\gamma$ = 2.86622 × $10^{-8}$ W$^{-1}$, and $\eta$ = 3.6× $10^{-8}$ W$^{-2}$. On the other hand, the optical parameters of the calculated FML pulses are with $\lambda_0$ = 1562.1 nm, $\Delta\lambda$ = 13.6 nm, $P_i$ = 1.84 mW, $P_o$ = 47.32 mW, $G$ = 14.11 dB, $\alpha$ = 45.48 mW, and repetition rate $R$ = 3.5 MHz, leading to $E_p$ = 7.02 nJ, $\Omega$ = 10.5 THz, $g$ = 7.02, $\delta$ = 0.12 W$^{-1}$, $\gamma$ = 0.005 W$^{-1}$, and $\eta$ = 7.15× $10^{-6}$ W$^{-2}$. According to the spectral width, the coherent length of FML pulses in the fiber is estimated to be 0.120 mm, which is close to the spatial pulse length of 0.143 mm. However, the coherent length of NML pulses in the fiber is 1.12 mm, which is much smaller than the spatial pulse length of 15360 mm. Therefore, the effective SPM parameter of NML pulses in the 56-m cavity should be scaled down by a factor of about 50000, thus $\delta$ = 9.97 × $10^{-8}$. Based on these system parameters, the experimental data of pulsewidths 76.8 ns and 714 fs are marked by blue squares and circles, respectively.

For NML pulses, Fig. 4(a) shows that in the normal dispersion regime, the pulsewidth grows dramatically and shows the possibility to reach the microsecond pulses at strong normal dispersion. On the other hand, for several-tens-meter fiber laser system with anomalous dispersion, there were no experimental findings to support the possibility in the generation of stable NML pulses. However, our calculation shows that a broad NML pulse solution is available from Eq. (2) with $B > 0$. In addition to the weak pulse shortening strength under tiny positive cubic SA, the inclusion of positive quintic SA provides a further compression mechanism to enhance the pulse formation. In fact, by adjusting the polarization controller, we are able to adjust the magnitude of the SA parameters and to obtain the broad NML pulses. For FML pulses, Fig. 4(b) shows that the pulse broadening for both large normal and anomalous intracavity dispersion are nearly linear with comparable magnitude. However, there is some discrepancy between our calculations and Haus's work, in which the broadening of the pulsewidth is only remarkable in the system with normal dispersion and positive SPM. We found that the discrepancy comes from the inclusion of an external quintic nonlinear absorption and the shaping parameter $B$ introduced in the ansatz function of Eq. (2). For FML pulses, the system with positive SPM and $B$ has the potential to form broader pulses at anomalous dispersions but the broadening effect is limited by negative $B$ at the normal dispersions.

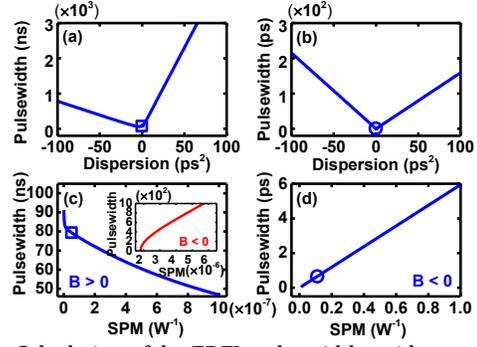

Fig. 4. Calculation of the EDFL pulsewidths with respect to (a)-(b) intracavity dispersions, and (c)-(d) SPM parameters.

In Fig. 4(c) and (d) we depict the pulsewidth as a function of SPM parameter. In a dispersive system, there exists a shortest Fourier-transform-limited pulsewidth for a specific SPM. However, our calculations give monotonic functions in both NML and FML cases. For NML pulses the optimal value of SPM is larger than the calculated SPM range, whereas it is smaller than the calculated range for FML pulses. Therefore, there shows opposite tendencies on the variation of the pulsewidth with SPM. Furthermore, the shaping parameter $B$ also verifies itself to play the role for assisting the pulse broadening or compression. We claim that while the minimum NML and FML pulsewidths would be restricted by the Fourier transform limit, the disregard of spatial dependence of the optical parameters restrict the calculation ranges in the current model. Therefore, we suggest that an optimal SPM can be obtained if the variation of bandwidth with SPM is considered. However, this implies that the analytical investigations of pulse formation mechanisms become no longer available, and the spatial dynamics of the pulse energy and associated generation of new frequencies have to be considered in the pulse evolution process. Within the mean-field model, our calculation provides a qualitative but successful interpretation on the NML and FML pulses generation that are observed in the experiments. Furthermore, we also found that there exists a second solution with even larger pulsewidth for NML pulses in larger SPM regime, as shown in the inset of Fig. 4(c).

In Fig. 5 we discuss the influences of SA on the pulsewidth. For conventional SA materials, the transmission has a monotonic dependence of the incident intensity and can be approximately simulated in terms of a cubic SA parameter. The SA plays the role of the temporal aperture that permits high intensity transmission. On the contrary, for artificial NPE SA, the relation between transmission and intensity is found to be a slowly-decaying sinusoidal function. This directly implies that for a better and consistent illustration of our experimental results, the inclusion of additional mechanism beyond the cubic interaction is crucial. In Fig. 5(a) and (b) we illustrate the dependence of pulsewidth on the cubic SA parameter $\chi$, where the pulsewidth varies inversely with SA for both NML and FML pulses. The experimental data of pulsewidths 76.8 ns and 714 fs are marked by blue square and circle, respectively. In Fig. 5(c) and (d) we show the dependence of pulsewidth on the quintic SA parameter $\eta$. We found that the system requires positive $\eta$ and shaping parameter $B$ to support the formation of stable NML pulses, while negative $\eta$ and $B$ is obtained for this cavity to form the FML pulses. The distinction can be attributed to the differences between the transmission of material SA and NPE SA. In low intensity regimes, the transmission curves of the two SA mechanisms show opposite sign of curvatures, indicating the existence of a highly nonlinear interaction to enhance SA for NPE at low intensities. The enhancement is available by using a positive $\eta$ for the NML pulses. However, for FML pulses a negative $\eta$ gives rise to the reduction of SA at high intensity regimes.

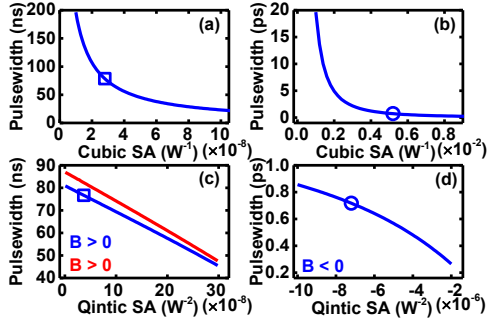

Fig. 5. Calculation of the EDFL pulsewidths with respect to (a)-(b) cubic SA parameters, and (c)-(d) quintic SA

The red curve in Fig. 5(c) shows that there exists a bistable state for the NML pulses whenever the quintic SA is considered, which also justifies the contribution of high-order nonlinear effect in the mean-field model. Moreover, Fig. 5 shows that the quintic SA is comparable with the cubic SA for the NML pulses, and is even larger under the current experimental setup. On the other hand, the cubic SA is larger than the quintic SA for FML pulses by several orders of magnitude. The violation of the nonlinear convergence is a direct disclosure of the difference between NPE SA and material SA. The transmission curve for NPE SA is concave upwards at low power intensities, whereas it is concave downwards for material SA. It turns out that the high-order nonlinear effects get to be crucial at low power intensities for NPE SA.

For this laser cavity, fundamental mode-locking with Kelly side bands can be observed. The laser output power is increased to 10.24 mW, and the pulse repetition rate is slightly increased to 3.67 MHz. While operating in this state, the optical spectrum has a center wavelength of 1559.2 nm and a 3-dB bandwidth of 4.0 nm. Using an autocorrelator, the pulsewidth is measured to be 1.89 ps. The adjustment of the polarization controller also enables us to adjust the envelope width of noise-like pulses from hundreds of picoseconds to several nanoseconds. In one of the adjustments, the envelope width is measured to be 184 ps, and the width of noise-like pulses is 473 fs. Therefore, a passively mode-locked fiber laser with a pulsewidth tuning range larger than five orders has been achieved by intracavity polarization adjustment.

In conclusion, we have constructed a ring-cavity passively mode-locked EDFL whose pulsewidth can be switched from 473 fs to 76.8 ns. Various laser operation states, including fundamental mode-locking, noise-like pulse, nanosecond mode-locking, and dual-width mode-locking have been obtained by simply adjusting an intracavity polarization controller. The envelope width of noise-like pulses ranges from hundreds of picoseconds to several nanoseconds. For dual-width mode-locking, the power ratio of broad and short pulses is 1:1.04. Therefore, the co-existence of femto/nano-second pulses is attributed to the balancing of net gain for the two lasing modes. We present an analytic modeling of the mode-locked fiber laser based on the cubic-quintic Ginzburg-Landau equation, in which the pulsewidths are calculated as functions of intracavity dispersion, saturable absorption, and self-phase modulation. Our calculation shows that a broad NML pulse solution is available in the CQGLE modeling with positive quintic SA and positive shaping parameter. The generation of nanosecond pulses in this EDFL is attributed to the weakened intracavity pulse-shortening strength and reduced effective SPM parameter in the laser cavity. The mode-locked laser pulses with temporal duration ranging over five orders may find useful applications in the future.

**Funding.** Ministry of Science and Technology (MOST), Taiwan (MOST 106-2221-E-845-002, MOST 104-2511-S-845-009-MY3)